\pdfoutput=1
\documentclass[12pt,preprint,epsf]{aastex} 

\usepackage{epstopdf}

\shorttitle{Spectrophotometry of Bright Stars}
\shortauthors{Krisciunas et al.} 

\begin{document}
\received{1 December 2016}

\title{Spectrophotometry of Very Bright Stars in the Southern Sky}   
\author{
Kevin Krisciunas,\altaffilmark{1,2}
Nicholas B. Suntzeff,\altaffilmark{1,2}
Bethany Kelarek,\altaffilmark{1}
Kyle Bonar,\altaffilmark{1} and
Joshua Stenzel\altaffilmark{1}
}
\altaffiltext{1}{Texas A. \& M. University, Department of Physics \& Astronomy,
  4242 TAMU, College Station, TX 77843; {krisciunas@physics.tamu.edu} }

\altaffiltext{2}{George P. and Cynthia Woods Mitchell Institute for Fundamental 
Physics \& Astronomy }

\begin{abstract} 

We obtained spectra of 26 bright stars in the southern sky, including Sirius, 
Canopus, Betelgeuse, Rigel, Bellatrix, and Procyon, using the 1.5-m telescope 
at Cerro Tololo Inter-American Observatory and its grating spectrograph 
RCSPEC.  A 7.5 magnitude neutral density filter was used to keep from saturating
the CCD. Our spectra are tied to a Kurucz model of Sirius with T = 9850 
K, log $g$ = 4.30, and [Fe/H] =+0.4.  Since Sirius is much less problematic 
than using Vega as a fundamental calibrator, the synthetic photometry of our 
stars constitutes a Sirius-based system that could be used as a new anchor 
for stellar and extragalactic photometric measurements.

\end{abstract}


\section{Introduction}

Flux calibration, whether it be for photometry or spectroscopy, is a 
fundamental aspect of observational astronomy \citep{Hea96,Hea14}. Vega 
($\alpha$ Lyr) has been a fundamental photometric and spectroscopic standard 
for decades \citep{Hay_Lat75, Boh14, Boh_etal14}. In the 1980's the {\em 
Infrared Astronomy Satellite} (IRAS) discovered circumstellar material around 
Vega.  Subsequently, observations with the {\em Spitzer Space Telescope} 
characterized this material as a debris disk \citep{Su_etal05, Su_etal13}. 
Vega may be spectroscopically variable as well \citep{But14}. 
\citet{Boh14} comments on the {\em non}-variability of Vega. In any case 
Vega is problematic as a fundamental calibrator.

The Sloan Digital Sky Survey committed to using four principal photometric 
standards \citep{Fuk_etal96}, but in the end relied primarly on the star BD 
+17\arcdeg 4708.  Recently, it was revealed that this star brightened by 0.04 
mag in the $UBVR$ bands from 1986 to 1991 \citep{Boh_Lan15}.

Here we present a set of bright spectrophotometric standards, many of the 
brightest stars visible in the southern hemisphere during the southern summer. 
Our data expand the lists of stars observed by \citet{Ham_etal92,Ham_etal94} 
and \citet{Str_etal05}. Given the increase in sensitivity of instrumentation 
over the years, it might be the first time in 40 years that carefully 
calibrated spectra of most of these bright stars have been obtained.  Using well 
defined bandpasses \citep{Bes90}, we can use our spectra to generate $BVRI$ 
photometry tied to a model of Sirius, which is a ``well behaved'' star 
compared to Vega.

\section{The Target Stars}

The target stars are situated from $-$70\arcdeg ~to +9\arcdeg ~declination, 
and all but one have right ascensions ranging from $\sim$5 to 13 hours (see 
Fig. \ref{fig:positions} and Table \ref{tab:stars}).  About half the target 
stars are members of binary or multiple star systems. $\alpha$ CMa (Sirius) 
and $\alpha$ Car (Canopus) are the two brightest stars in the night sky. 
$\zeta$ Ori is the brightest O-type star in the sky.

Two other notable stars are $\epsilon$ CMa and $\beta$ CMa.  The former was
the brightest star in the sky 4.7 million years ago (with visual
magnitude -3.99).  The latter was the brightest star in the sky 4.4 million
years ago (with visual magnitude -3.65).  This was not due to changes in
their {\em intrinsic} luminosities.  It was due to their changing {\em distances}
from the Sun \citep{Tom98}.

$\alpha$ Ori (Betelgeuse) is a variable star.  On the basis of 17 years of 
photoelectric photometry by one of us (KK), we found that its $V$-band 
magnitude ranges from 0.27 to 1.00 \citep{Kri82, Kri90}. A photoelectric light 
curve obtained from 1979 through 1996 is shown in Fig. \ref{fig:aori}.  The 
mean brightness during these years was $V$ = 0.58.

On the basis of all sky photometry and differential photometry 
with respect to $\phi ^2$ Ori, it appears that HR 1790 ($\gamma$ Ori; Bellatrix)
ranges in brightness in the $V$-band by as much as 0.07 mag \citep{Kri_Fis88}.

Only three of our targets ($\beta$ Ori, $\alpha$ CMa, and $\alpha$ CMi) are
fundamental $UBV$ standards of \citet{Joh_Mor53}.  Their targets are primarily
northern hemisphere objects.

\section{Data Acquisition and Reduction}

Three nights were allocated to this project on the CTIO 1.5-m telescope in
January of 2003, and eight more nights were allocated in January of 2005.
However, due to a variety of hardware and weather programs, we only 
obtained useful data on two nights, 6 January 2003 and 21 January 2005 (UT).
On the first night all the spectra were taken with the blue grating.  On
the second night all the spectra were taken with the red grating.

Details of the facility spectrograph RCSPEC are discussed by \citet{Str_etal05}. 
The blue and red gratings give dispersions of 2.85 and 5.43 \AA ~per pixel, 
respectively.  \citet{Str_etal05} give 5.34 \AA ~per pixel as the dispersion
of the red grating, but this is a transcription error.
The FWHM values are 8.6 \AA ~for the blue grating and 
16.4 \AA ~for the red grating. Because of the extreme brightness most of our 
stars, we included a 7.5 mag neutral density filter in the light path to prevent 
saturation of the pixels.  Our exposure times ranged from 5 to 420 seconds.  
While the spectra of Landolt standards obtained by \citet{Str_etal05} are useful 
at wavelengths as short as 3100 \AA, ours are no good below 3300 \AA.

Raw two dimensional spectra were saved as FITS files 1274 by 140 pixels in size.  
Batches of four spectra were taken of each star, with the telescope offset 30 
arcsec west along the slit between spectra to place the spectrum on a different part of the 
chip.  A He-Ar-Ne arc spectrum was taken before every batch.\footnote[3]{For 
calibration line identification we used A CCD Atlas of Helium/Neon/Argon Spectra, 
by E. Carder, which can be downloaded at https://www.noao.edu/kpno/KPManuals/henear.pdf.} 
Once the star was centered in a  2\arcsec ~slit, the slit width was widened to 
21\arcsec.  Since many of our targets are close binary or multiple stars, 
this means that many of our spectra  are blended spectra of more than one star.  
On the plus side, such a wide slit eliminates any worries about guiding and 
seeing, allowing accurate spectrophotometry under clear sky conditions.

Spectra were reduced in the {\sc iraf} environment.\footnote[4]{{\sc iraf} 
is distributed by the National Optical Astronomy Observatory,
which is operated by the Association of Universities for Research in
Astronomy, Inc., under cooperative agreement with the National Science
Foundation (NSF).}  We made extensive use of the spectroscopic reduction
manual of \citet{Mas_etal92}.
We first bias subtracted, trimmed, and flattened the spectra.
One dimensional spectra were extracted with {\em apall} in the {\sc apextract}
package.\footnote[5]{{\sc iraf} users should know or be reminded that there is
a second version of {\em apall} in the {\sc ctioslit} part of {\sc imred}.
Setting the many parameters in one version of {\em apall} does not set 
them in the other parameter list!}

Wavelength calibration was accomplished using {\em identify}, {\em reidentify}, 
and {\em dispcor} in the {\sc ctioslit} package.  Once we had carried out the 
wavelength calibration we could ask the question: To what extent were our two 
useable nights clear?  To do this one can sum up the instrumental counts over 
some wavelength range, then take $-$2.5 log$_{10}$ of the counts to produce 
instrumental magnitudes. In Fig. \ref{fig:clear_sky} we show these instrumental 
magnitudes vs. airmass from 27 Sirius spectra obtained on 6 January 2003. We have 
eliminated the 9 spectra that were the final spectra of the batches of four on 
this date.  For reasons we do not understand the final spectrum of each batch 
often gave an instrumental magnitude that was about $\sim$0.10 mag fainter than 
the other three.  In Fig. \ref{fig:clear_sky} the slope is 0.237 $\pm$ 0.009 mag 
per airmass.  The RMS residual of the fit is $\pm$0.018 mag, which is comparable 
to CCD photometry on a photometric night. The wavelength range for integrating 
those spectra was 3600 to 5500 \AA.  This is somewhat wider than the standard 
$B$-band filter.  From photometry at Cerro Tololo and Las Campanas we find a mean 
$B$-band extinction coefficient of 0.262 $\pm$ 0.007 mag per airmass.  The bottom line is 
that by using RCSPEC as a photometer, we demonstrated that 6 January 2003 was 
clear the whole night.

Similar considerations for the spectra taken on 21 January 2005 indicate that 
this night became non-photometric by 05:27 UT.  We will only consider 
spectra taken on this night prior to this time.

The flux calibration of our spectra was carried out with tasks {\em standard}, 
{\em sensfunc}, and {\em calibrate} within the {\sc ctioslit} package.  With 
the {\em calibrate} task we applied extinction corrections appropriate for
Cerro Tololo (found in file onedstds\$ctioextinct.dat within {\sc iraf}). 

For flux calibration of the blue grating spectra obtained on 6 January 2003 we observed
the spectrophotometric standards HR 3454 (observed at a mean airmass of 1.201) and HR 4468 (observed 
at mean airmass 1.154). For red grating spectra obtained on 21 January 2005 we used 
the standard HR 1544 for the flux calibration.  It was observed at a mean airmass of 1.427. The 
mean airmass values for the observations of our target stars are given in Table 
\ref{tab:stars}.  Any systematic errors in the flux calibration with {\sc iraf} 
will be equal to the arithmetic {\em difference} of the airmass of the standards 
and the program stars {\em times} the arithmetic {\em difference} of the true extinction 
coefficient as a function of wavelength minus the adopted mean values appropriate
to CTIO. For photometric sky and observations above an 
elevation angle of 45 degrees, any systematic error of the flux calibration should 
be less than 10 percent in the $B$-band and less than 5 percent in the $VRI$ bands.
{\em Relative} fluxes of our spectra and synthetic photometry have estimated internal
random errors of 3 percent or better (see below).

The final step in our reduction was to tie the spectra to a Kurucz model of 
Sirius. An {\sc ascii} version of an R=1000 spectrum of Sirius was kindly 
provided by Ralph Bohlin.\footnote[6]{One must use a model spectrum of 
appropriate resolution.  Otherwise the final spectra may contain spurious 
features such as fictitious P Cygni profiles. A scaled FITS version of the 
Kurucz model spectrum can be obtained via 
http://www.stsci.edu/hst/observatory/crds/calspec.html as file 
sirius\_mod\_002.fits.  A comment in the header of this file indicates that 
fluxes have been scaled by 2.75440 $\times 10^{-16}$. This accounts for the 
distance to Sirius and its limb-darkened angular diameter.}
The sampling is at twice the frequency of the resolution. The model spectrum 
has T = 9850 K, log $g$ = 4.30 and metallicity [Fe/H] = +0.4.

The wavelengths of the model spectrum were in nm, so we multiplied by 10 to 
convert them to \AA.  We also want wavelengths in air, rather than vacuum 
wavelengths.  For this we used the transformation given at the SDSS Data 
Release 7 website.\footnote[7]{ 
http://classic.sdss.org/dr7/products/spectra/vacwavelength.html} Finally, we 
used a scale factor of 2.75440 $\times 10^{-16}$ to convert the
Kurucz model flux to that of Sirius, so that it is measured in erg cm$^{-2}$ s$^{-1}$ \AA$^{-1}$.

In the top panel of Fig. \ref{fig:sirius} we see the Kurucz model spectrum. 
The middle panel is the average of 18 blue grating spectra of Sirius (taken at 
airmass less than 1.3), and 16 red grating spectra, as processed with {\sc 
iraf}.  We have stitched together the blue grating spectra and the red grating 
spectra at 6000 \AA, which produces a small discontinuity at that wavelength.  
The bottom panel of Fig. \ref{fig:sirius} is the {\em ratio} of the Kurucz 
spectrum and our mean Sirius spectrum.  We call that ratio the ``flux 
function'' or ``spectral flat''.  All our other spectra are then multiplied by 
the flux function to place them on a system tied to the Kurucz model of 
Sirius.  This largely, but not entirely, takes out the discontinuity at 6000 \AA 
~and also takes out telluric features such as the Fraunhofer B-, A-, and Z-lines 
at 6867, 7594 and 8227 \AA, which are due to atmospheric O$_2$. The identity of
a feature at $\sim$3680 \AA ~evident in many of our spectra is uncertain;
it too is largely taken out by the spectral flat.

The average value of the flux function shown in the bottom panel of Fig. 
\ref{fig:sirius} is 0.990, which is close enough to 1.000 to give us confidence 
that the flux calibration of our coadded spectra of Sirius, obtained with {\sc 
iraf}, is consistent with the scaling of the model of Sirius to the flux 
density of the star. Our ultimate filter by filter zeropoints are the values of 
the $BVRI$ magnitudes of Sirius given in Table \ref{tab:stars}, which come from 
\citet{Cou71,Cou80}.


Fully reduced spectra, transformed to the ``Sirius system'' and ranging from
3300 to 10,000 \AA, are shown in Fig. \ref{fig:bluered}.  Spectra taken
with only the blue grating are shown in Fig.~\ref{fig:blue}.\footnote[8]{
FITS and ASCII spectra are available via
http://people.physics.tamu.edu/krisciunas/spec.tar.gz and from the online version
of this paper.}

Some line identifications are given in the last panel of Fig. \ref{fig:bluered}.
In spectra of stars hotter than the Sun we clearly see the Balmer lines at 6563, 4861, 4340, 
4102 \AA\ and shorter wavelengths.  Cooler stars such as HR 3307 ($\epsilon$ Car)
and HR 2061 ($\alpha$ Ori) show the infrared Ca$^+$ triplet (8498, 8542, and
8662 \AA) and the blended Na D lines (5890 and 5896 \AA). $\zeta$ Ori and early 
B-type stars, such as HR 1790 ($\gamma$ Ori), HR 2294 ($\beta$ CMa),
HR 2618 ($\epsilon$ CMa), and HR 4853 ($\beta$ Cru), show He I absorption at 
4471 and 5876 \AA, though it is difficult to see given the scale of the
spectra shown in Figs. \ref{fig:bluered} and \ref{fig:blue}.  
A higher resolution spectrum of $\zeta$ Ori A from 3980 to 4940 \AA, 
including line identifications, is shown in Fig. 14 of \citet{Sot_etal11}. 

One thing to note in our reduced spectra is the strength of the Balmer jump in 
early-type main sequence stars. This is due to ionization of atomic hydrogen 
from the first excited state, producing strong absorption shortward of the 
Balmer limit at 3646 \AA.  This results in fainter $U$-band magnitudes of such 
stars.  A much weaker Balmer jump is seen in hot giant and supergiant stars.  
Thus, the Balmer jump gives us a photometric tool to measure a combination of
the luminosity class and the local acceleration of gravity of hot stars (log $g$).  
For example, an A2 V star is 
0.30 mag redder in the $U-B$ color index than an A2 III star \citep[][pp. 
388-389]{Dri_Lan00}. \citet{Kal62} points out that one also needs the rotation 
rates of the stars to do this properly.

\section{Synthetic Photometry}

The filter prescriptions originally given by \citet{Bes90} have been slightly 
modified by \citet{Bes_Mur12}.  We have adopted the latter.  In Fig. \ref{fig:ubvri}
we show their filter prescriptions, multiplied by an atmospheric extinction function
appropriate to Cerro Tololo, and also multiplied by a function which accounts
for the principal atmospheric extinction lines. This is noticeable in the 
$R$- and $I$-band functions.

We then calculated synthetic $BVRI$ magnitudes of our target stars using 
an {\sc iraf} script written by one of us (N. B. S.).  This script uses an arbitrary 
zero point for each filter. We adjusted
the $BVRI$ zero points to given synthetic magnitudes of the scaled Sirius model
spectrum that match those of \citet{Cou71, Cou80}.  If the reader chooses to adopt
different $BVRI$ magnitudes of Sirius than those given in Table \ref{tab:stars},
then the synthetic magnitudes of the other stars given in the table must be adjusted
up or down accordingly.

Bessell and Murphy's $V$-band filter prescription extends to 7400 \AA, while our 
blue grating spectra stop at $\sim$6400 \AA. We cannot obtain synthetic 
$V$-band magnitudes for the cooler stars observed only with the blue grating.  
However, we can obtain approximate $V$-band magnitudes for the hot stars HR 2618, 
3485, 4853, and 4963 by extrapolating the spectra using the Rayleigh-Jeans approximation.

Table \ref{tab:stars} gives our synthetic $BVRI$ photometry. Fig. 
\ref{fig:pmdiff} shows the differences of our synthetic photometry and the 
values of \citet{Cou71} and \citet{Cou80}, as a function $B-V$ (for $B$ and
$V$), $V-R$ (for $R$), and $V-I$ (for $I$).  There is no color term  for the
$V$-band differences, but there are non-zero colors terms for $B$, $R$, and $I$.
At zero color there is no offset between our $V$-band magnitudes and those of
Cousins, but in $B$, $R$, and $I$ ours are 0.02 to 0.03 mag fainter.

%

From the AAVSO online light curve calculator we find that the $V$-band brightness 
of $\alpha$ Ori was $V$ = 0.398 on 2 January 2003, and $V$ = 0.384 on 7 January. 
The mean is $V$ = 0.391, which is comparable to our synthetic $V$-band 
magnitude of 0.398 from spectra taken on 6 January 2003.  This is a good sanity check.
On 21 January 2005, when we took the red grating spectra, Betelgeuse's brightness was 
$V$ = 0.436, according to the AAVSO.

The spectra presented here and the associated synthetic photometry can function as
a Sirius-based anchor for Galactic as well as extragalactic observational 
astronomy.

\acknowledgments

We made use of the SIMBAD database, operated at CDS, Strasbourg, France.  We 
thank the AAVSO for the $V$-band photometry of Betelgeuse obtained from their 
database.  Kenneth Luedeke and Raymond Thompson measured Betelgeuse closest 
to the times we took spectra.  We thank Ralph Bohlin for providing an ASCII 
version of the Kurucz spectrum of Sirius used for the calibration, and for 
useful comments.  We also thank James Kaler and Jesus Ma\'{i}z Apell\'{a}niz
for comments and references.

\appendix

\section{Other Spectra}

The spectra shown in Figs. \ref{fig:bluered} and \ref{fig:blue} were taken under demonstrably 
clear sky conditions. Synthetic photometry based on these spectra is transformable to
the systems of \citet{Cou71} and \citet{Cou80} with uncertainties of $\pm$0.03 mag 
or less.  Other spectra were taken which might be of use to the reader.

In Fig. \ref{fig:blue4} we show blue grating spectra of HR 5056 ($\alpha$ Vir) and
HR 5267 ($\beta$ Cen) taken on 6 January 2003.  For reasons that are not entirely clear,
our synthetic photometry was too faint by $\sim$0.55 mag and $\sim$0.12 mag for these two
stars.  The most likely explanation is a misalignment of the telescope and the
dome slit.  We have scaled these two spectra by appropriate amounts to
make them consistent with photometry of \citet{Cou71}.

In Fig. \ref{fig:red1} we show red grating spectra of HR 3454 ($\eta$ Hya), HR 4216
($\mu$ Vel), and HR 4450 ($\xi$ Hya), taken through clouds on 21 January 2005.
The spectra have been scaled to make them consistent with photometry of \citet{Cou80}.

Finally, in Fig. \ref{fig:etacar} we show two spectra of $\eta$ Carinae taken through
clouds on 21 January 2005. The top spectrum is a coadd of 12 exposures of 7 seconds.
Such a short exposure time was necessary to prevent saturation of the H-$\alpha$ line.
The bottom spectrum is a coadd of 3 exposures of 240 seconds.  In this spectrum
H-$\alpha$ is saturated, but other emission lines such as the Paschen lines of
hydrogen and multiple helium lines are evident with a better signal-to-noise ratio.
Since $\eta$ Car has such a non-stellar spectrum and we have no available $R$-
or $I$-band photometry of this star at this epoch, we have not scaled our spectra
like the others presented in this Appendix.

These additional spectra are available from the first author of this article.

\clearpage

\begin{deluxetable}{clclcccccc}
\rotate
\tablecolumns{10}
\tablewidth{0pc}
\tabletypesize{\scriptsize}
\tablecaption{Synthetic Photometry of Target Stars\label{tab:stars}}
\tablehead{ \colhead{HR\tablenotemark{a}} &
\colhead{Name} &
\colhead{Binary/Multiple?} & 
\colhead{Spectral Type\tablenotemark{b}} &
\colhead{X$_{blue}$\tablenotemark{c}} & 
\colhead{X$_{red}$\tablenotemark{d}} & 
\colhead{$B$} & 
\colhead{$V$} & 
\colhead{$R$} & 
\colhead{$I$} }
\startdata
1544                       &  $\pi ^2$ Ori    & N & A1Vn                  & 1.338 & 1.427 & \phn4.381 & \phn4.365  & \phn4.365 & \phn4.351 \\
1713                       &  $\beta$ Ori     & N & B8 Ia:                & 1.085 & 1.144 & \phn0.200 & \phn0.224  & \phn0.173 & \phn0.148 \\
1790                       &  $\gamma$ Ori    & N & B2 III                & 1.253 & 1.352 & \phn1.425 & \phn1.635  & \phn1.756 & \phn1.888 \\
1948/1949                  &  $\zeta$ Ori A/B & Y & O9.7b+O9 III+B0 II-IV & 1.376 & 1.223 & \phn1.635 & \phn1.826  & \phn1.874 & \phn1.968 \\
2061\tablenotemark{e}      &  $\alpha$ Ori    & N & M1-2 Ia-Iab           & 1.291 & 1.321 & \phn2.198 & \phn0.398  &  $-$0.653 &  $-$1.799 \\
2294                       &  $\beta$ CMa     & N & B1 II-III             & 1.316 & 1.123 & \phn1.761 & \phn1.996  & \phn2.126 & \phn2.265 \\
2326                       &  $\alpha$ Car    & N & F0 II                 & 1.251 & 1.137 &  $-$0.584 &  $-$0.726  &  $-$0.830 &  $-$0.952 \\
2491\tablenotemark{f}      &  $\alpha$ CMa    & Y & A1 V                  & 1.123 & 1.077 &  $-$1.425 &  $-$1.420  &  $-$1.408 &  $-$1.400 \\ 
2943                       &  $\alpha$ CMi    & Y & F5 IV-V               & 1.299 & 1.226 & \phn0.761 & \phn0.357  & \phn0.132 &  $-$0.098 \\
3307                       &  $\epsilon$ Car  & Y & K3 III+B2: V          & 1.211 & 1.147 & \phn3.081 & \phn1.844  & \phn1.092 & \phn0.360 \\
3685                       &  $\beta$ Car     & N & A2 IV                 & 1.428 & 1.305 & \phn1.674 & \phn1.661  & \phn1.693 & \phn1.690 \\
\hline
  99                       &  $\alpha$ Phe    & Y & K0 III                & 1.562 & \ldots & \phn3.460 &  \ldots    & \ldots & \ldots \\
2618                       &  $\epsilon$ CMa  & Y & B2 II                 & 1.226 & \ldots & \phn1.315 & \phn1.519  & \ldots & \ldots \\
2693                       &  $\delta$ CMa    & N & F8 Ia                 & 1.172 & \ldots & \phn2.533 &  \ldots    & \ldots & \ldots \\
3485                       &  $\delta$ Vel    & Y & A1 V                  & 1.177 & \ldots & \phn1.990 & \phn1.949  & \ldots & \ldots \\
3634                       &  $\lambda$ Vel   & N & K4.5 Ib-II            & 1.126 & \ldots & \phn3.822 &  \ldots    & \ldots & \ldots \\
3748                       &  $\alpha$ Hya    & N & K3 II-III             & 1.126 & \ldots & \phn3.415 &  \ldots    & \ldots & \ldots \\
4763                       &  $\gamma$ Cru    & N & M3.5 III              & 1.192 & \ldots & \phn3.160 &  \ldots    & \ldots & \ldots \\
4853                       &  $\beta$ Cru     & Y & B0.5 III              & 1.169 & \ldots & \phn1.050 & \phn1.241  & \ldots & \ldots \\
4963                       &  $\theta$ Vir    & Y & A1 IV s+Am            & 1.225 & \ldots & \phn4.392 & \phn4.364  & \ldots & \ldots \\
\enddata
\tablecomments{The stars listed in the top half of the table were observed with the blue grating and the
red grating.  The stars in the bottom half of the table were only observed with the blue grating.}
\tablenotetext{a}{Harvard Revised number = catalog number in {\em The Bright Star Catalogue}.}
\tablenotetext{b}{From online version of {\em The Bright Star Catalogue}, 5th edition, 1991.}
\tablenotetext{c}{Mean airmass for blue grating spectra obtained on 6 January 2003 UT.} 
\tablenotetext{d}{Mean airmass for red grating spectra obtained on 21 January 2005 UT.} 
\tablenotetext{e}{$\alpha$ Ori is variable.  See text for comments.}
\tablenotetext{f}{$BV$ photometry from \citet{Cou71}.  $RI$ photometry from \citet{Cou80}.}
\end{deluxetable}

\clearpage

\figcaption[]
{Positions of our target stars on the sky. The numerical labels are the catalog 
numbers in {\em The Bright Star Catalogue}. Blue dots represent stars that were observed
with the blue grating only.  Other stars were observed with both the blue and red
gratings.
\label{fig:positions}
}

\figcaption[]
{$V$-band magnitude of $\alpha$ Ori from October 1979 through November 1996.
Key to data points: blue dots (K. Krisciunas), green squares (D. Fisher),
red triangles (K. Luedeke).  Data by Fisher were published by \citet{Kri_Fis88}.
Data by Luedeke were published by \citet{Kri_Lue96}.
\label{fig:aori}
}

\figcaption[]
{Instrumental magnitudes of Sirius vs. airmass on 6 January 2003 (UT).  The
Y-axis values are equal to $-$2.5 log$_{10}$ of the integrated counts
from 3600 to 5500 \AA ~of spectra that have only been wavelength-calibrated.
\label{fig:clear_sky}
}

\figcaption[]
{{\em Top:} Kurucz model spectrum of Sirius with R = 1000, T = 9850 K, log $g$ 
= 4.30, and [Fe/H] = +0.4.  {\em Middle:} Average of calibrated Sirius spectra 
taken with airmass less than 1.3.  This is the output from {\sc iraf}.  {\em 
Bottom:} Ratio of model spectrum of Sirius to output from {\sc iraf}.  This is 
the ``flux function'' or ``spectral flat'' used to multiply the reduced 
spectra of the other stars to place the spectra on the system of the Kurucz 
model of Sirius.  In the bottom two panels the Fraunhofer B, A, and Z lines 
at 6867, 7594, and 8227 \AA\ are labeled.  These are due to molecular oxygen 
in the Earth's atmosphere.  
\label{fig:sirius}
}

\figcaption[]
{Spectra of program stars observed with the blue grating and the red grating.
Small spurious variations of the flux density are seen in some spectra 
at 6000 and $\sim$7600 \AA ~which are attributable to the method of construction
of the ``spectral flat''.  Some line identifications are given in the last panel
of this figure.  See text for further information.
\label{fig:bluered}
}

\figcaption[]
{Spectra of program stars observed with the blue grating only.
\label{fig:blue}
}

\figcaption[]
{Fractional transmission of \citet{Bes_Mur12} filters, multiplied by an atmospheric
extinction model appropriate to Cerro Tololo, and also multiplied by 
a function that accounts for principal terrestrial atmospheric features.
As our spectra are not useful shortward of 3300 \AA, designated here by
a vertical dashed line, we can not easily obtain $U$-band synthetic photometry 
of our program stars.
\label{fig:ubvri}
}

\figcaption[]
{Differences of photometry in the sense ``our synthetic photometry'' minus
``photometry of Cousins''.  The slope is also known as the ``color term''.
\label{fig:pmdiff}
}

\figcaption[]
{Spectra of HR 5056 ($\alpha$ Vir) and HR 5267 ($\beta$ Cen) obtained on
6 January 2003.
\label{fig:blue4}
}

\figcaption[]
{Spectra of HR 3454 ($\eta$ Hya), HR 4216 ($\mu$ Vel), and HR 4450
($\xi$ Hya)  obtained on 21 January 2005 under non-photometric conditions.
\label{fig:red1}
}

\figcaption[]
{Spectra of $\eta$ Carinae, obtained under non-photometric conditions on
21 January 2005.
{\em Top:} Coadd of shorter exposures.  The H-$\alpha$ line is not saturated.  
{\em Bottom:} Coadd of longer exposures. While the H-$\alpha$ line is
saturated, other spectral features are more easily seen. 
\label{fig:etacar}
}

\clearpage

\begin{figure}
\plotone{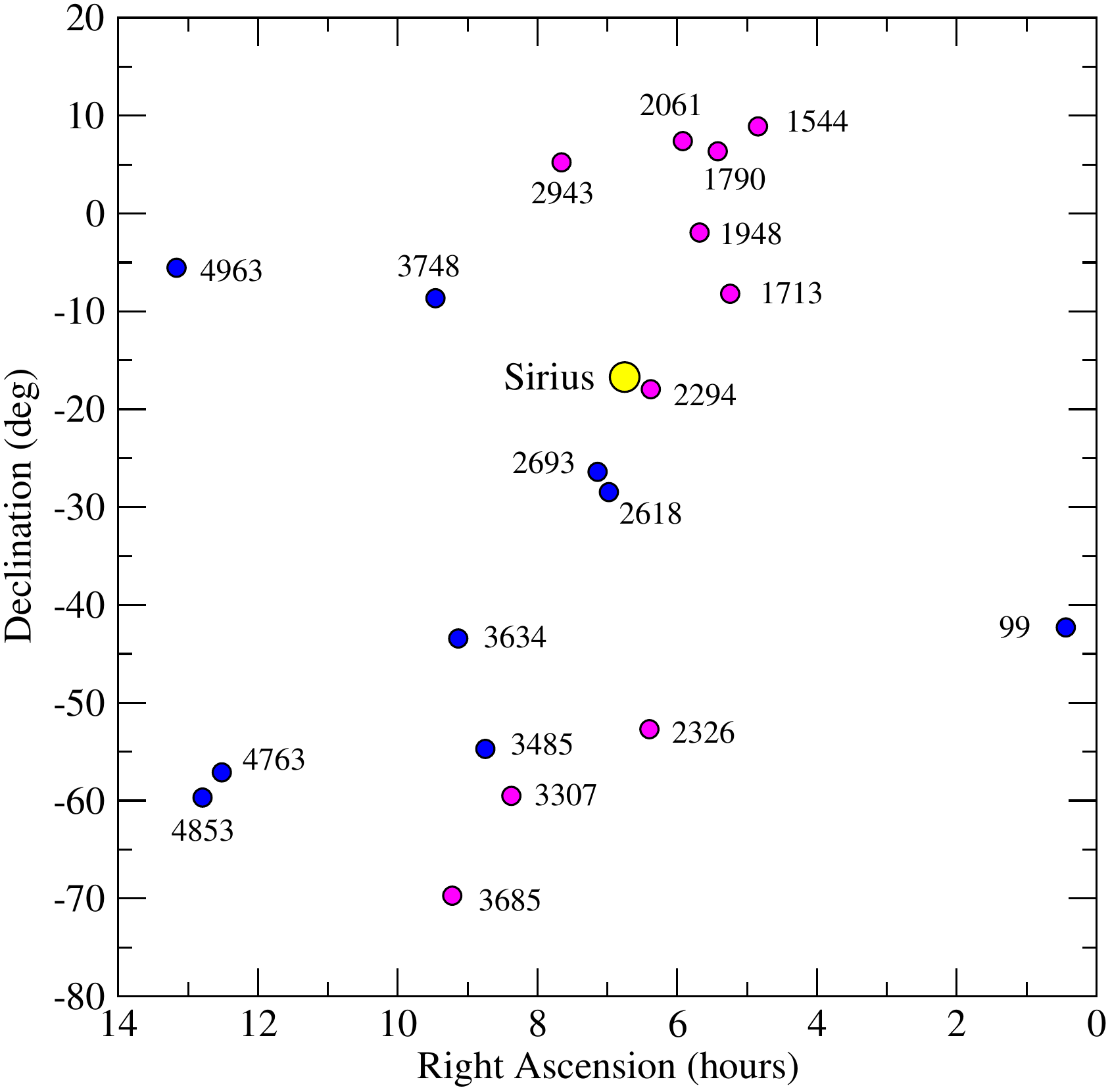} {Krisciunas Fig. \ref{fig:positions}. 
}
\end{figure}

\begin{figure}
\plotone{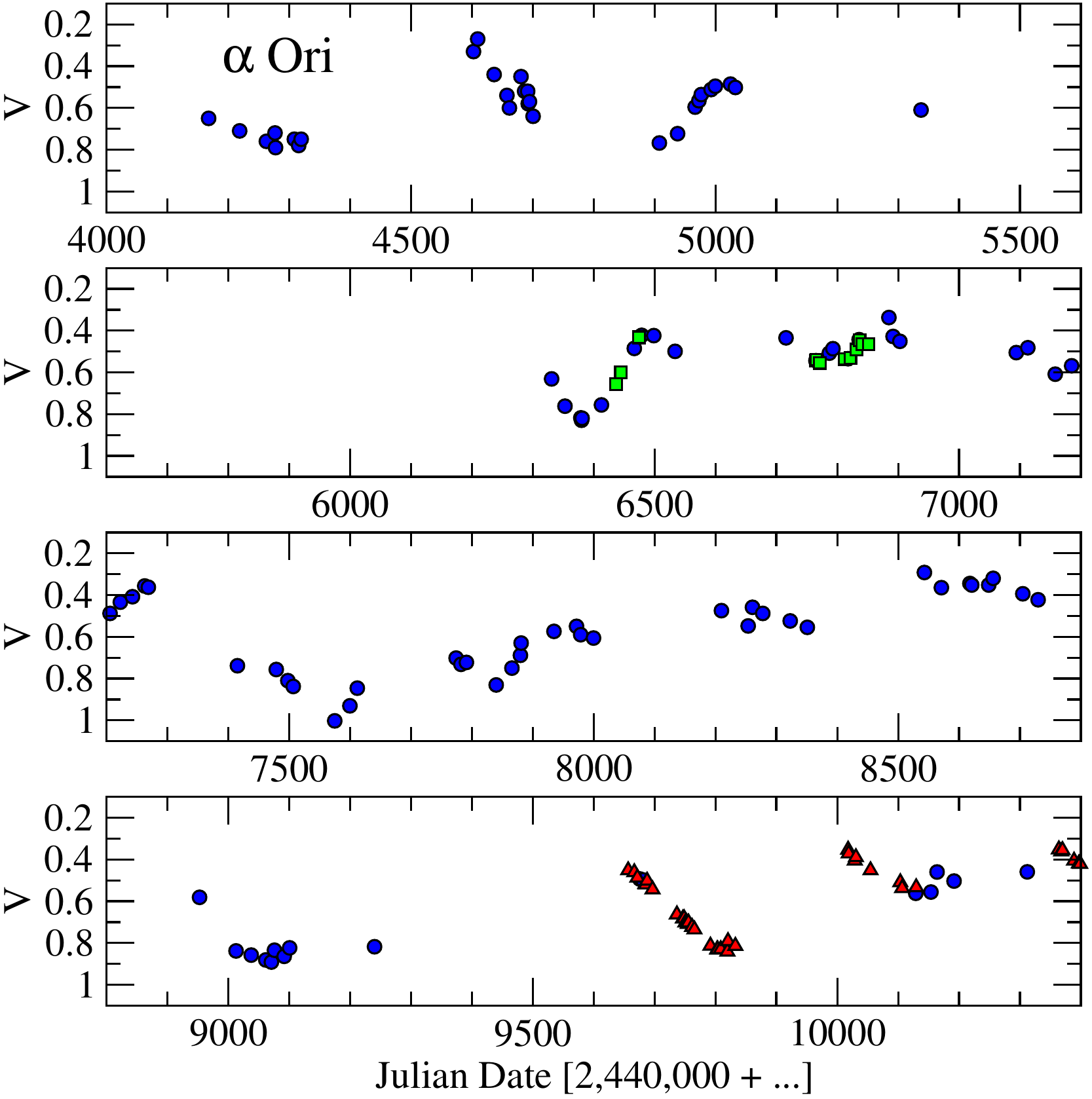} {Krisciunas Fig. \ref{fig:aori}. 
}
\end{figure}

\begin{figure}
\plotone{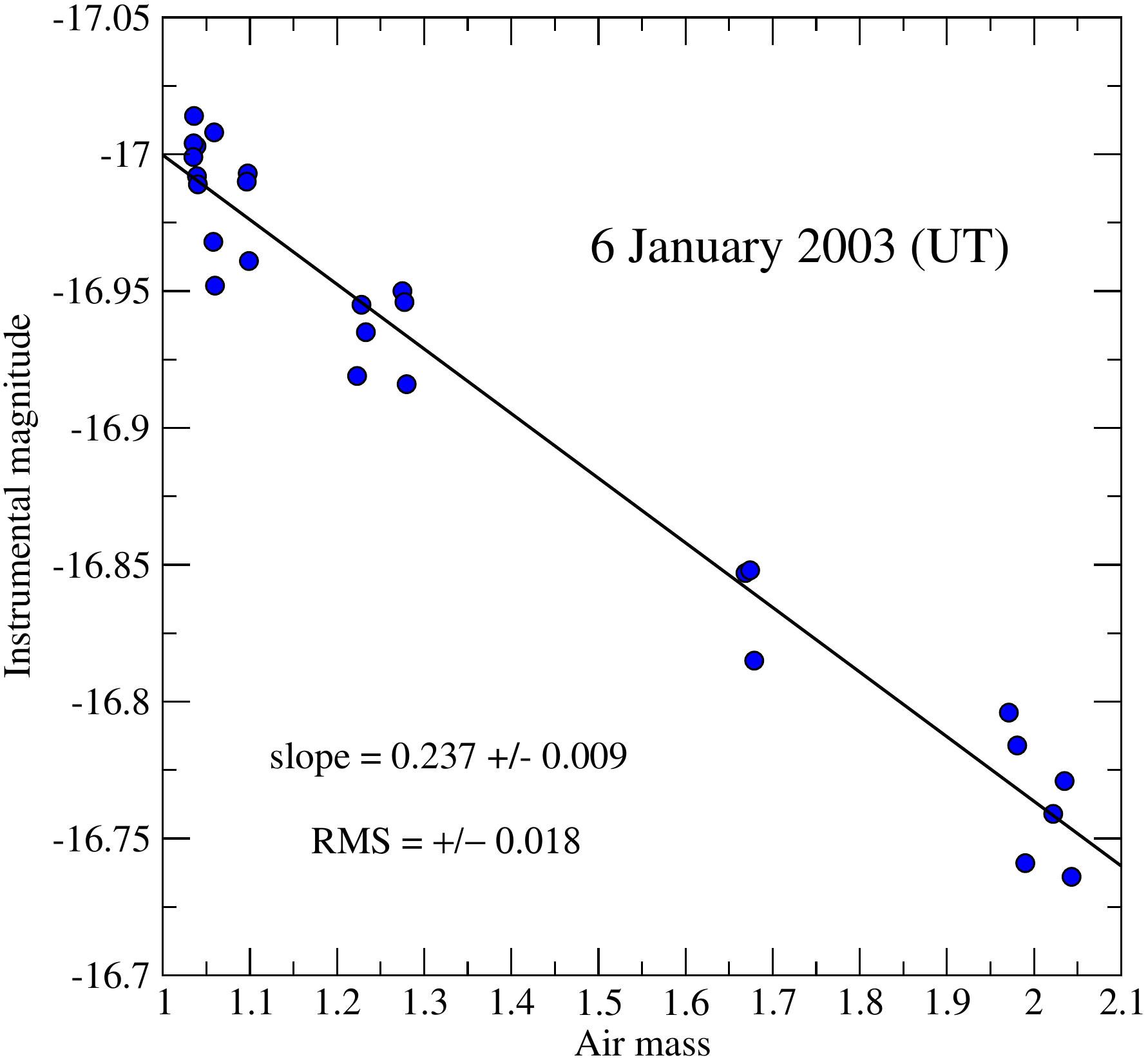} {Krisciunas Fig. \ref{fig:clear_sky}. 
}
\end{figure}

\begin{figure}
\plotone{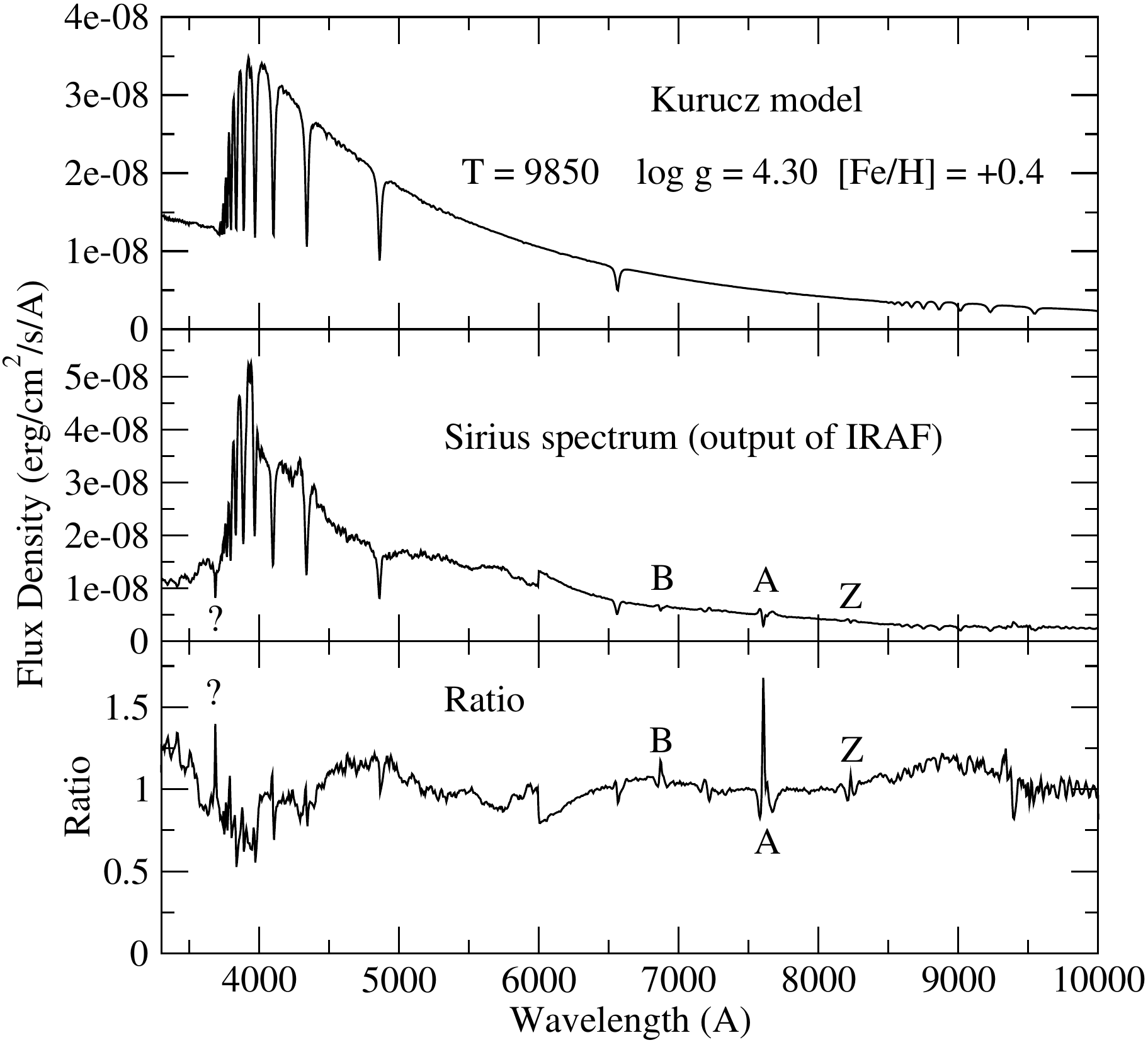} {Krisciunas Fig. \ref{fig:sirius}. 
}
\end{figure}

\begin{figure}[t]
\plotone{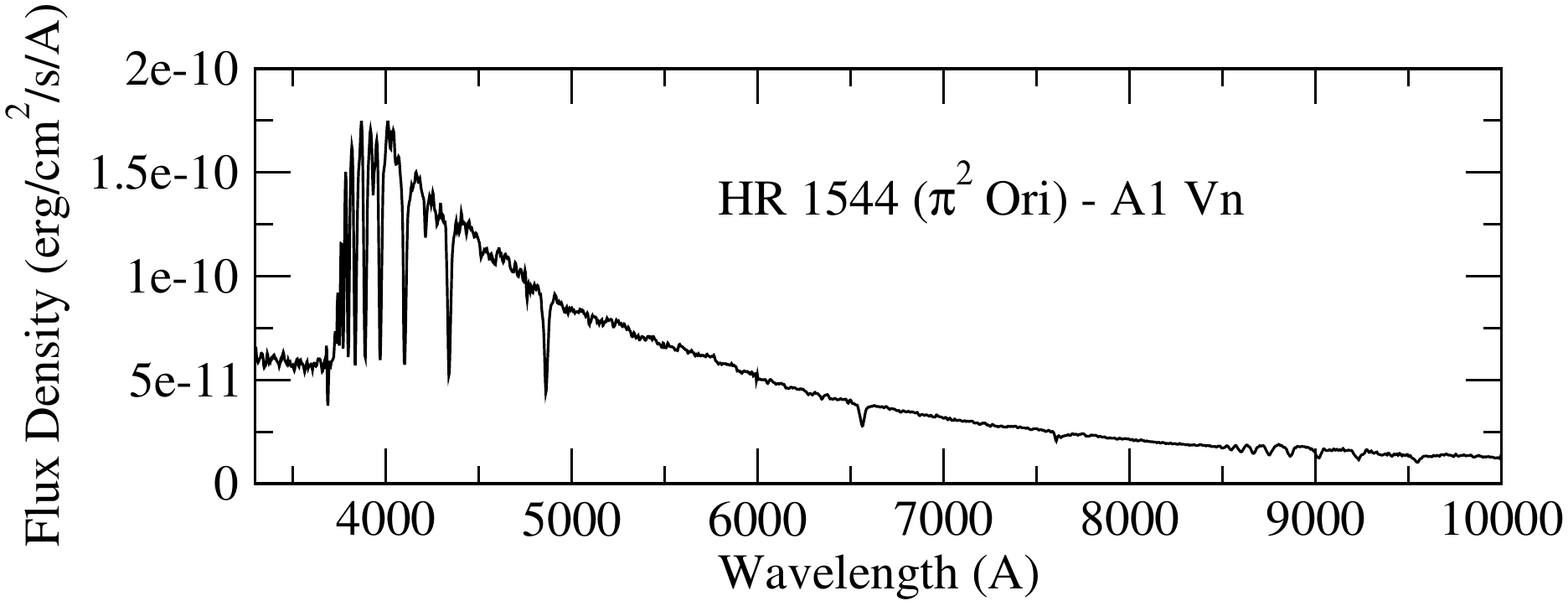}
{\center Krisciunas {\it et al.} Fig.~\ref{fig:bluered}}
\end{figure}
\clearpage
\newpage

\begin{figure}[t]
\plotone{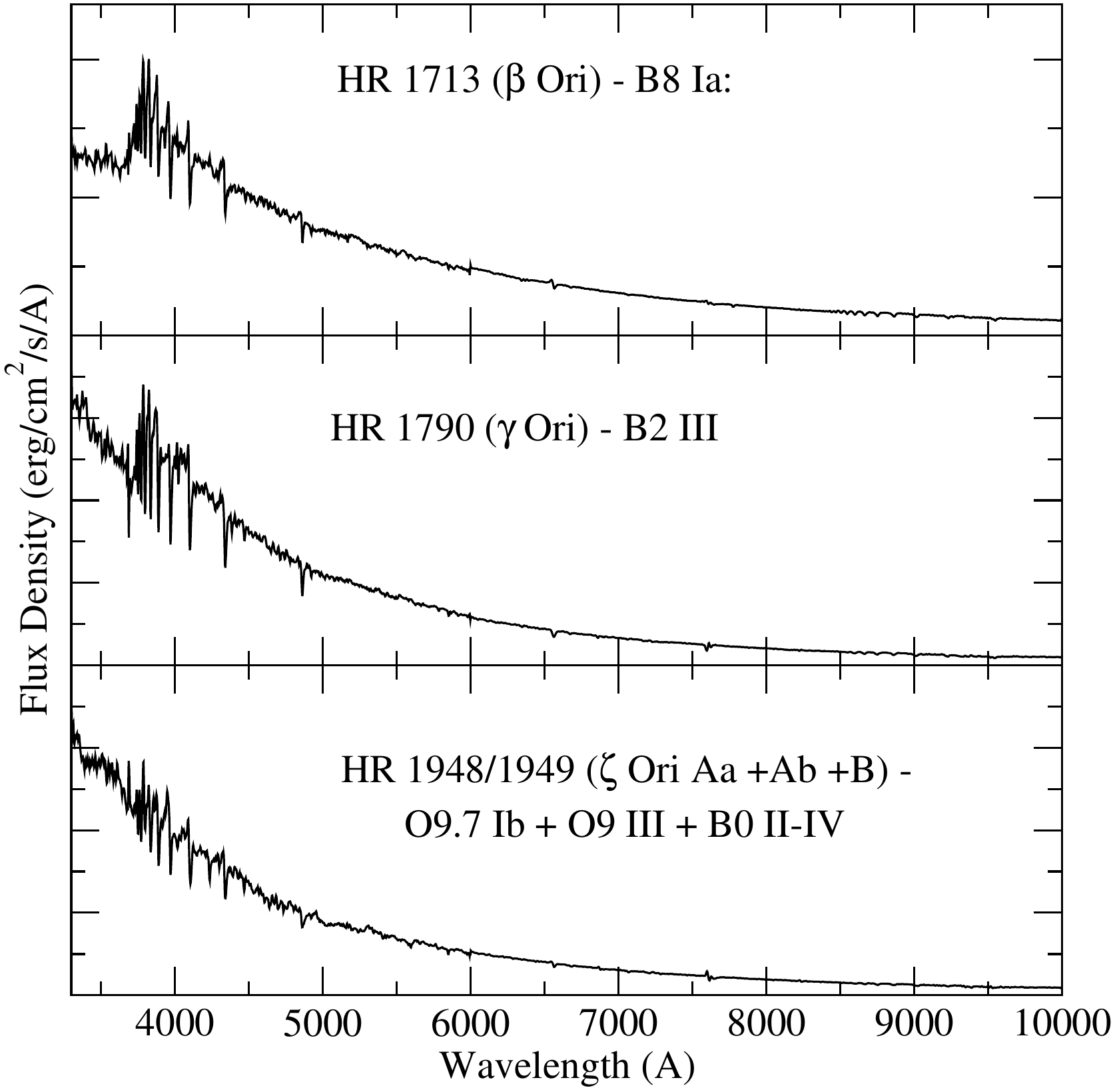}
{\center Krisciunas {\it et al.} Fig.~\ref{fig:bluered} (Continued)}
\end{figure}
\clearpage
\newpage

\begin{figure}[t]
\plotone{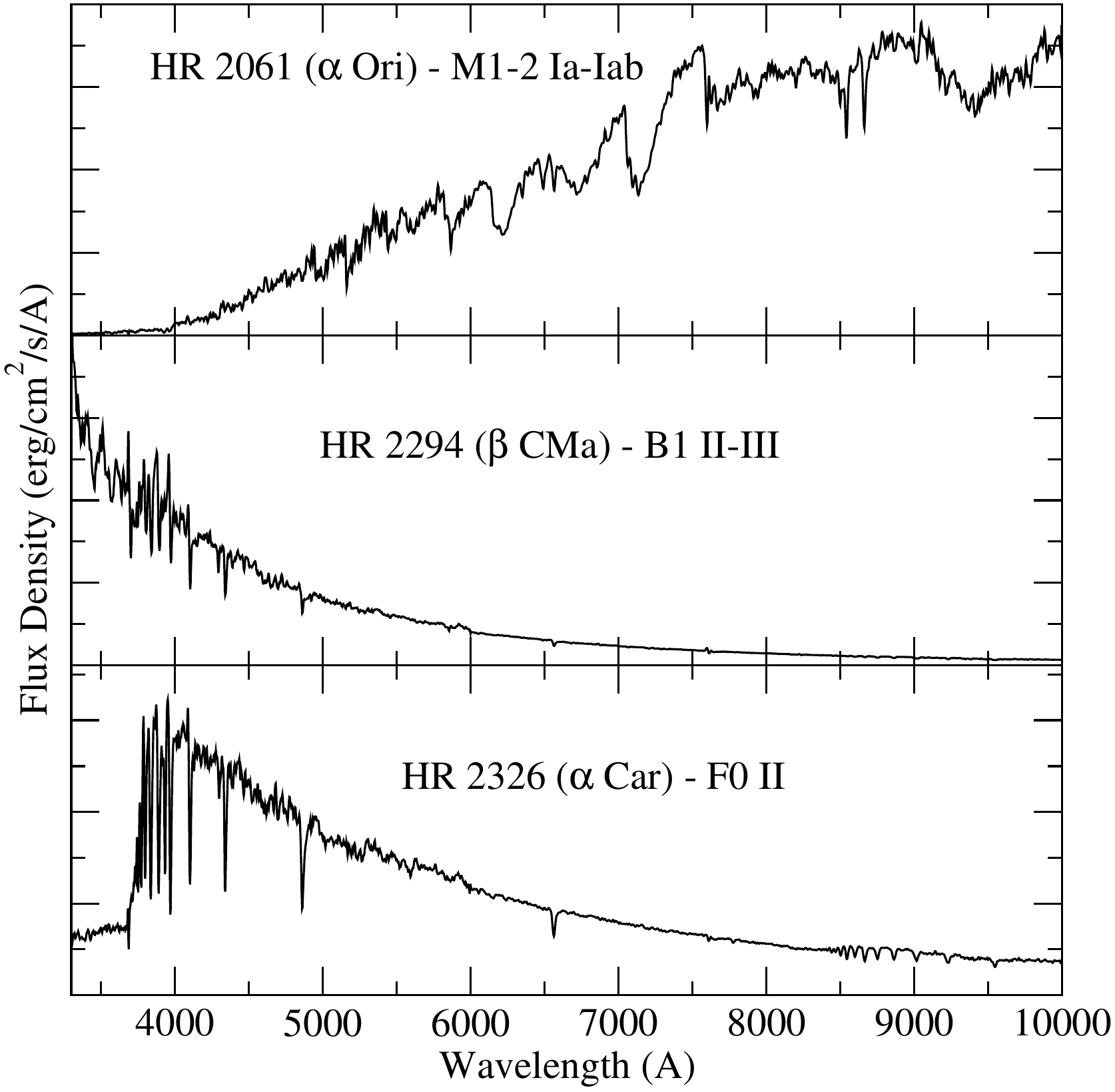}
{\center Krisciunas {\it et al.} Fig.~\ref{fig:bluered} (Continued)}
\end{figure}
\clearpage
\newpage

\begin{figure}[t]
\plotone{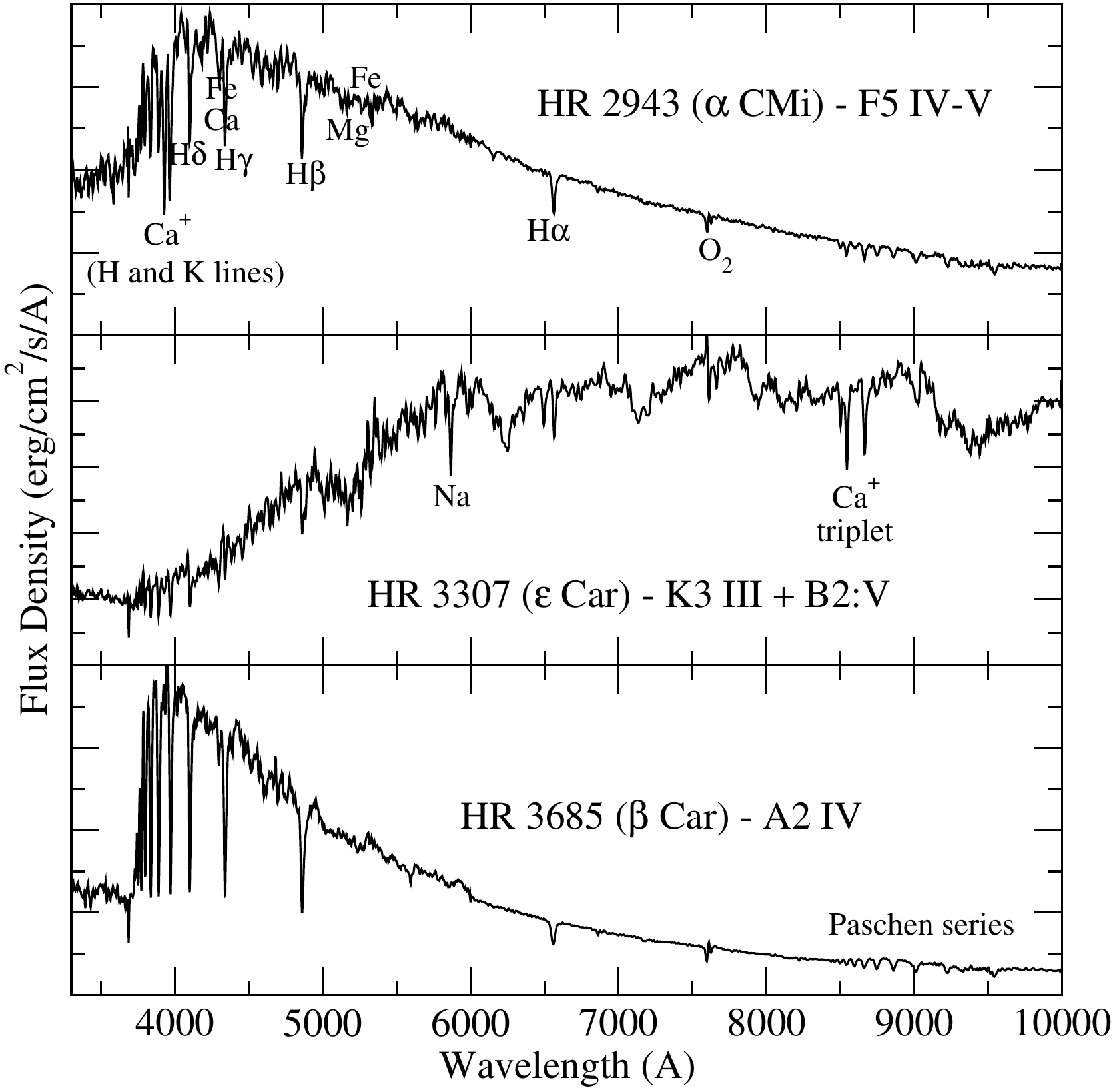}
{\center Krisciunas {\it et al.} Fig.~\ref{fig:bluered} (Continued)}
\end{figure}
\clearpage
\newpage

\begin{figure}[t]
\plotone{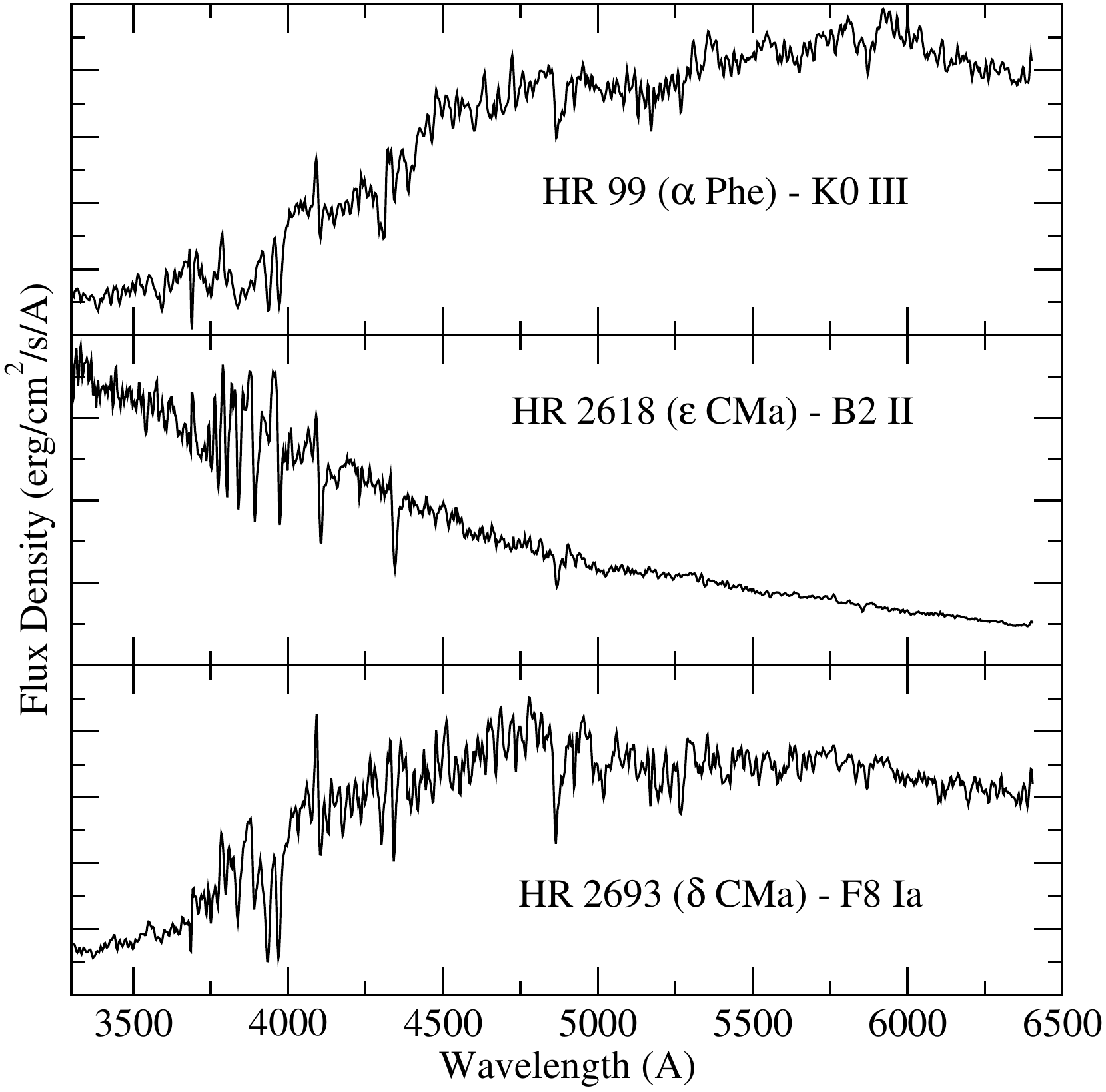}
{\center Krisciunas {\it et al.} Fig.~\ref{fig:blue}}
\end{figure}
\clearpage
\newpage

\begin{figure}[t]
\plotone{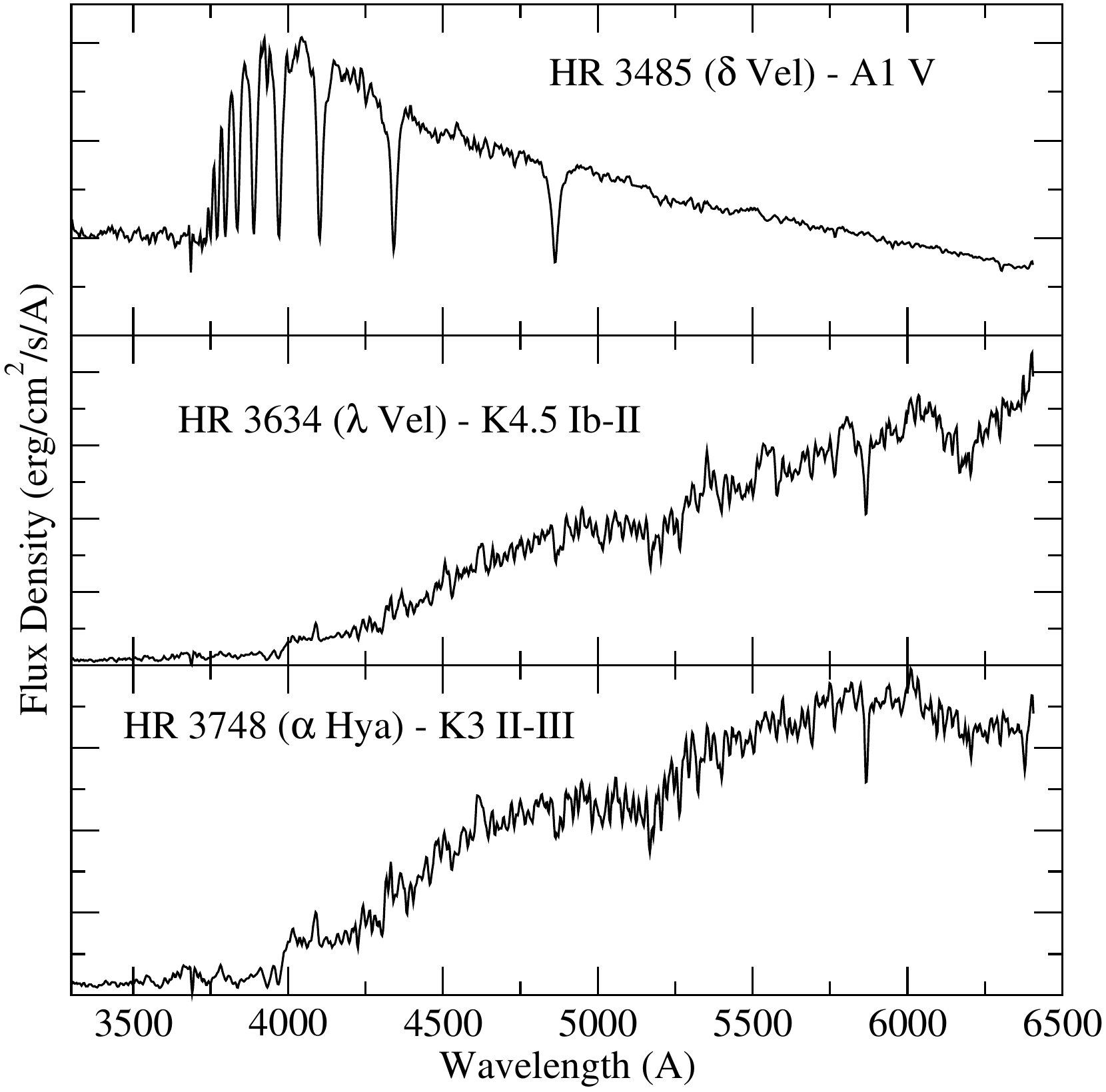}
{\center Krisciunas {\it et al.} Fig.~\ref{fig:blue} (Continued)}
\end{figure}
\clearpage
\newpage

\begin{figure}[t]
\plotone{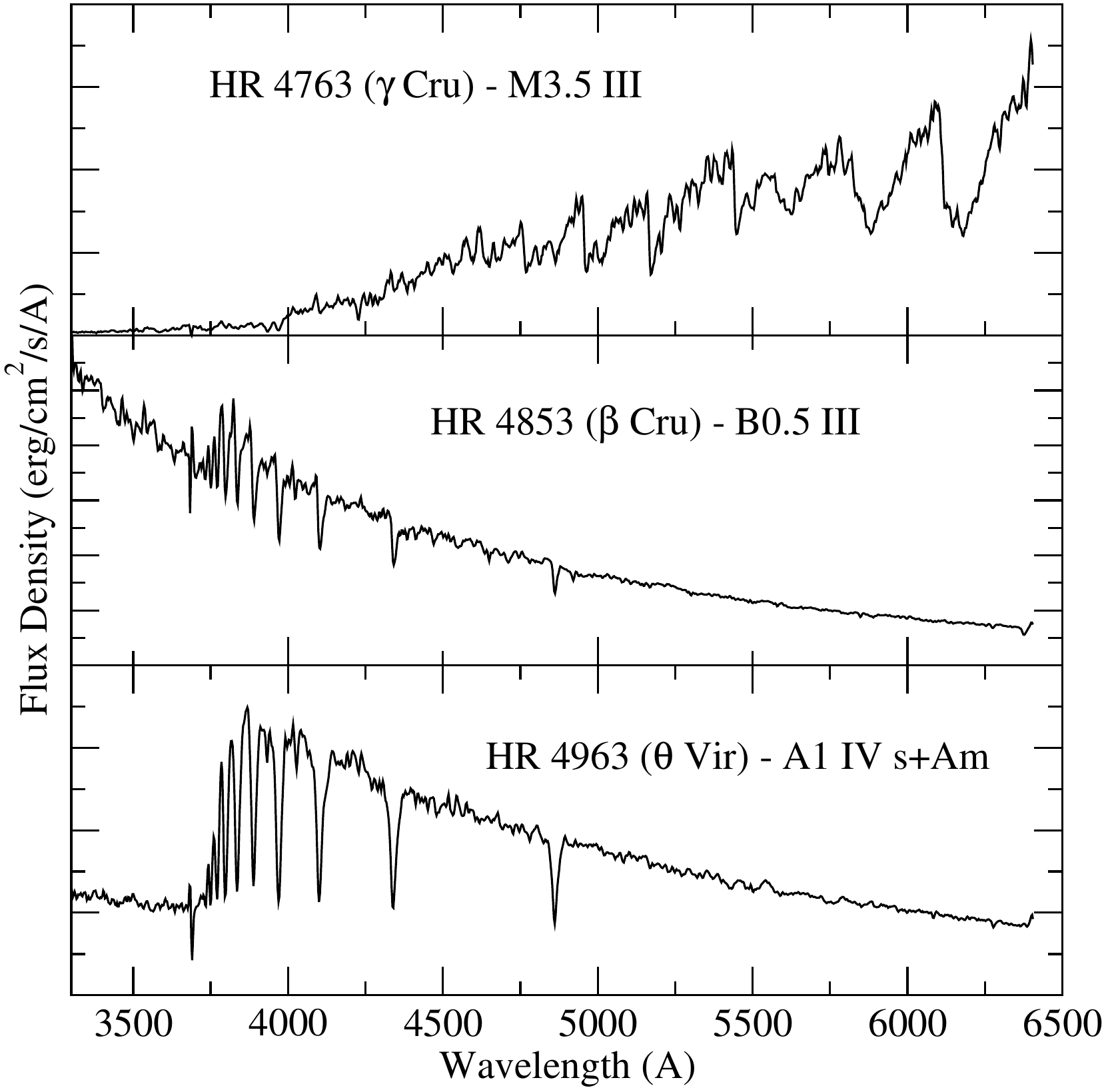}
{\center Krisciunas {\it et al.} Fig.~\ref{fig:blue} (Continued)}
\end{figure}
\clearpage
\newpage

\begin{figure}
\plotone{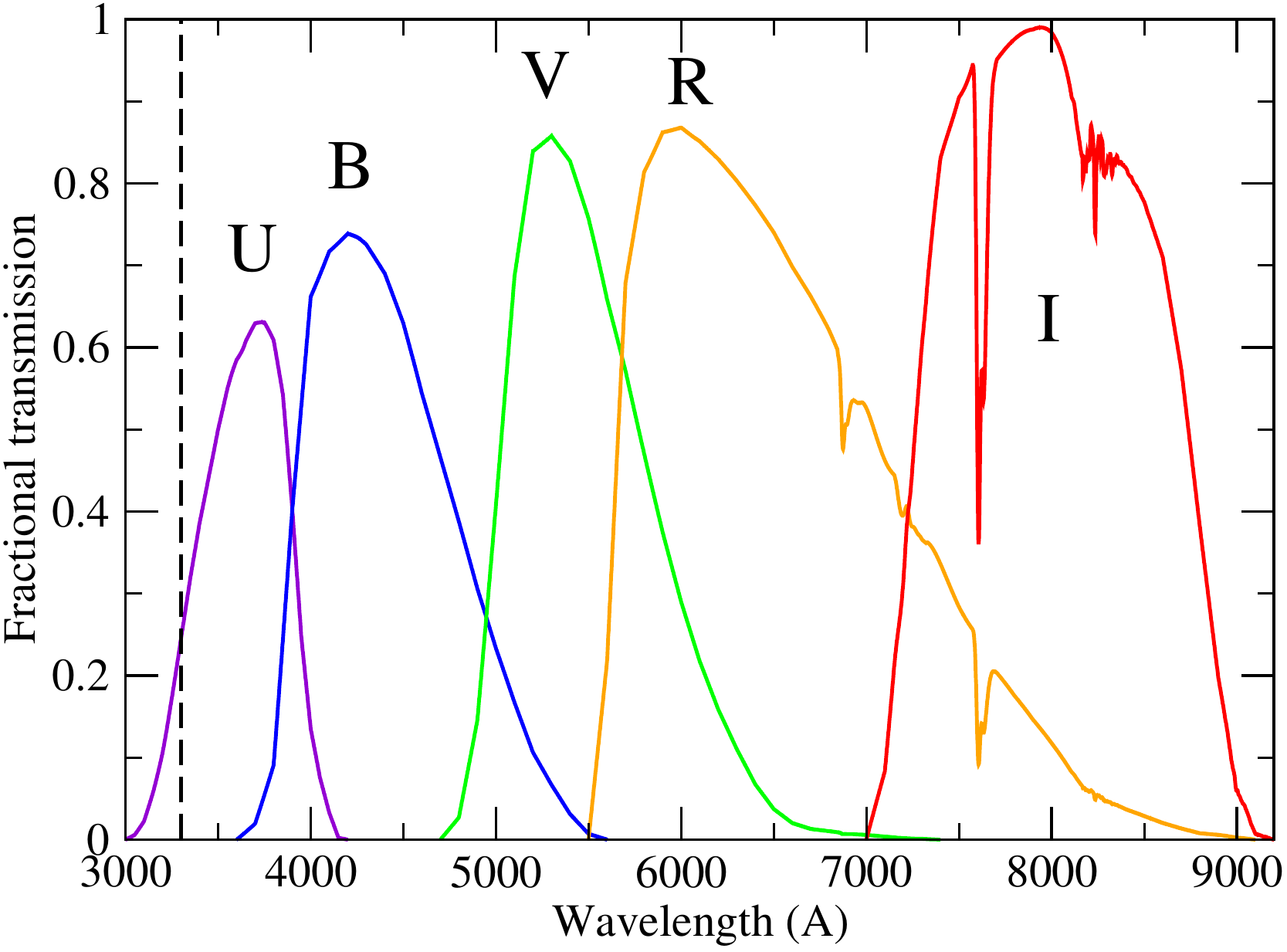} {Krisciunas Fig. \ref{fig:ubvri}. 
}
\end{figure}

\begin{figure}
\plotone{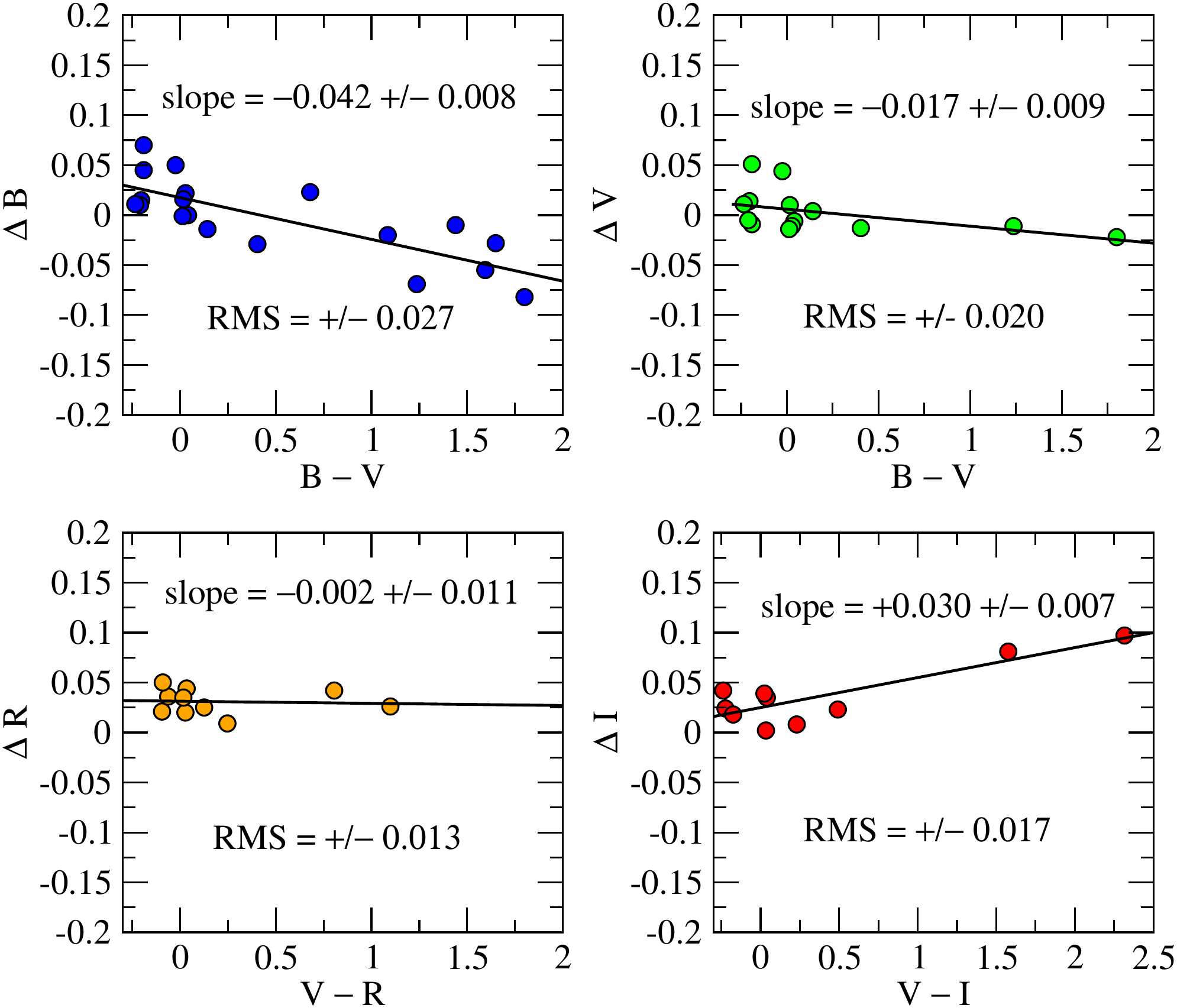} {Krisciunas Fig. \ref{fig:pmdiff}. 
}
\end{figure}

\begin{figure}
\plotone{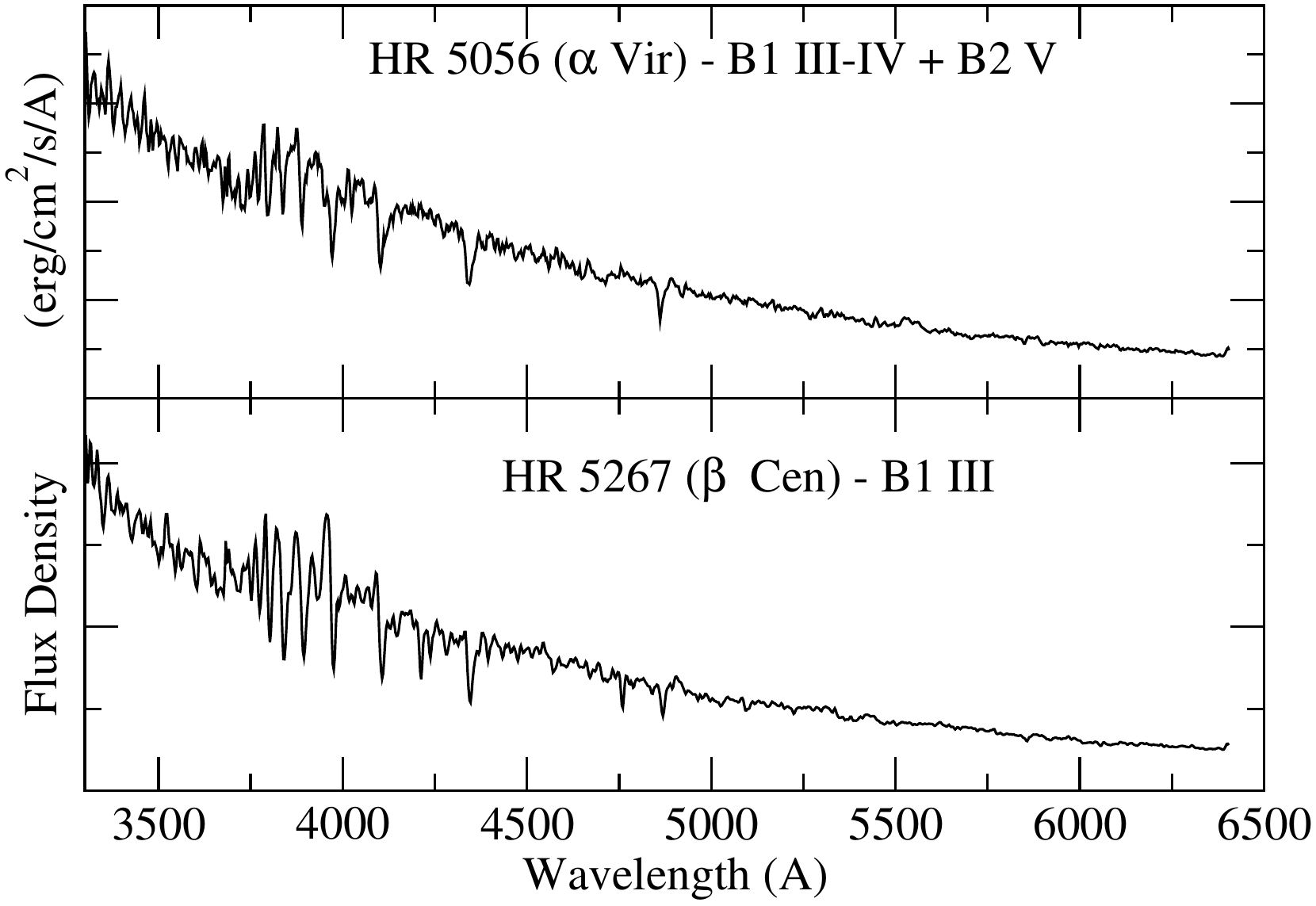} {Krisciunas Fig. \ref{fig:blue4}. 
}
\end{figure}

\begin{figure}
\plotone{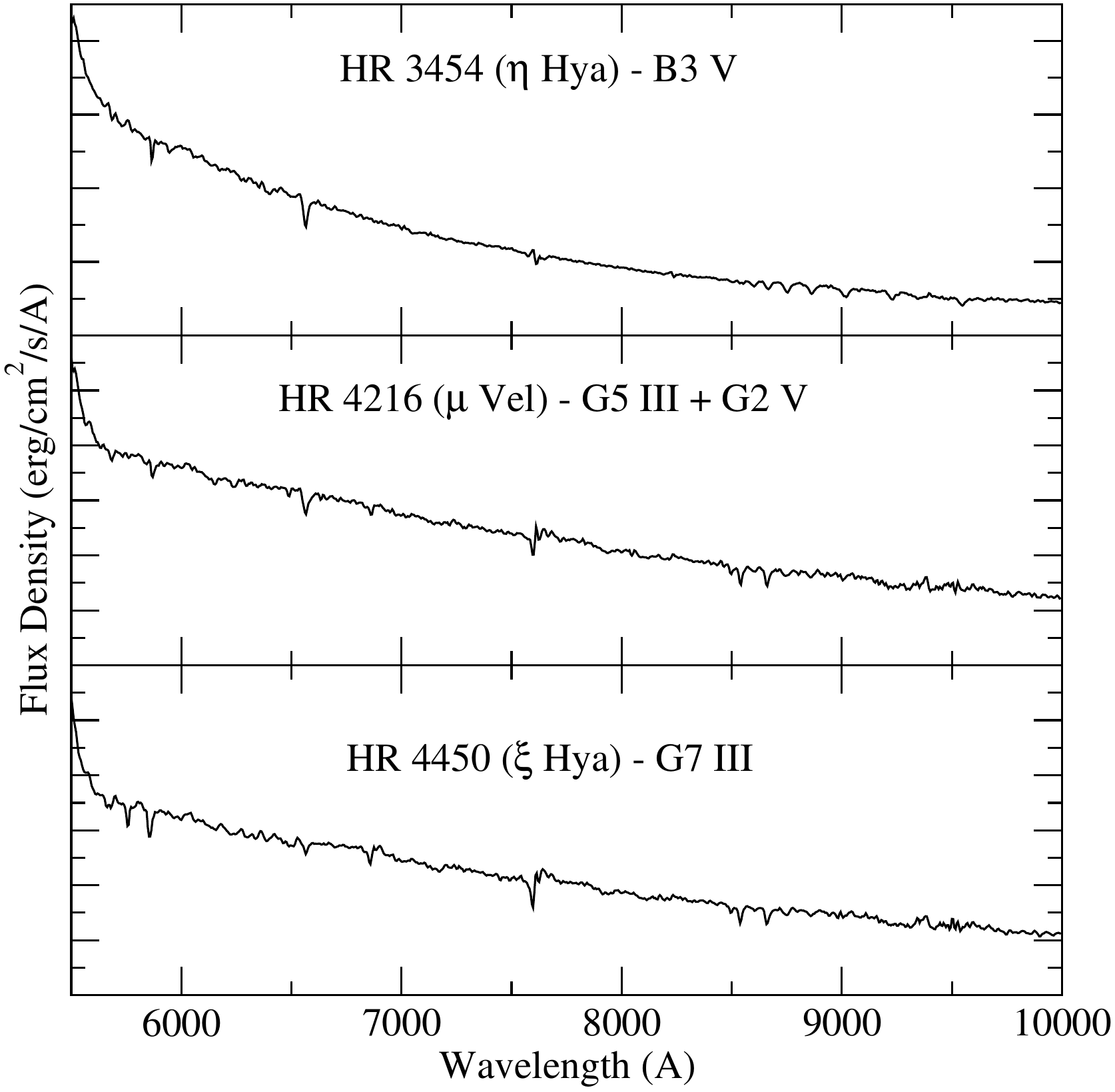} {Krisciunas Fig. \ref{fig:red1}. 
}
\end{figure}

\begin{figure}
\plotone{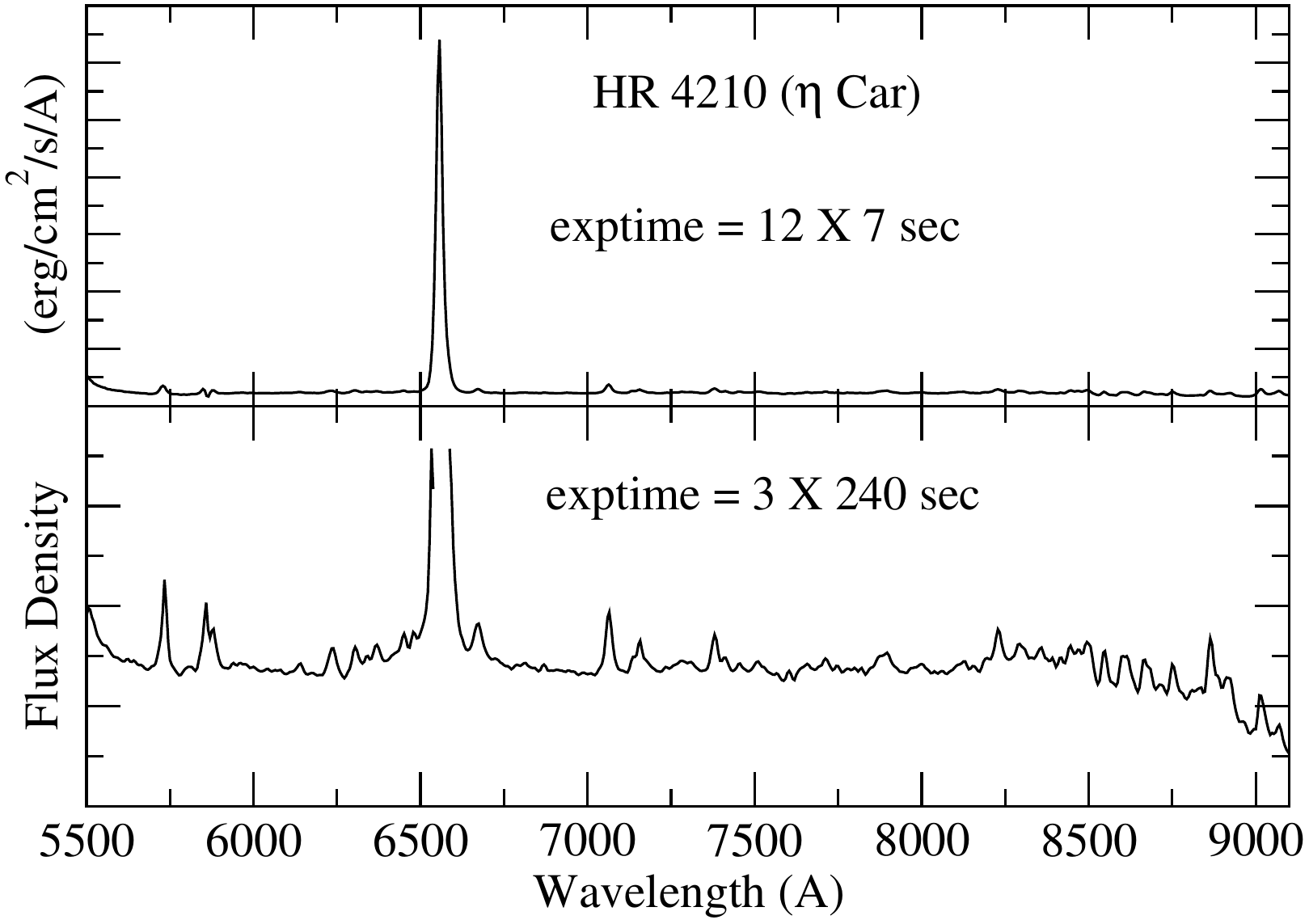} {Krisciunas Fig. \ref{fig:etacar}. 
}
\end{figure}

\end{document}